\documentstyle[aps,prd,epsf]{revtex}

\newcommand{\be}{\begin{equation}}
\newcommand{\ee}{\end{equation}}
\newcommand{\ben}{\begin{eqnarray}
\displaystyle}
\newcommand{\een}{\end{eqnarray}}

\newcommand{\p}{\partial}
\newcommand{\na}{\nabla}

\newcommand{\tV}{{\tilde V}}

\newcommand{\tg}{{\tilde g}}

\newcommand{\hhE}{{\hat E}}

\newcommand{\trho}{{\tilde \rho}}
\newcommand{\th}{{\tilde h}}

\newcommand{\tR}{{\tilde R}}

\newcommand{\ga}{\gamma}

\begin{document}

\title{Uniqueness Theorem of Static Degenerate and Non-degenerate 
Charged Black Holes in Higher Dimensions}

\author{Marek Rogatko}

\address{Institute of Physics \protect \\
Maria Curie-Sklodowska University \protect \\
20-031 Lublin, pl.Marii Curie-Sklodowskiej 1, Poland \protect \\
rogat@tytan.umcs.lublin.pl \protect \\
rogat@kft.umcs.lublin.pl}

\date{\today}

\maketitle
\begin{abstract}
We prove the uniqueness theorem for static higher dimensional charged black holes
spacetime
containing an asymptotically flat spacelike hypersurface with compact interior and
with both degenerate and non-degenerate components of the event horizon.
\end{abstract}

\pacs{04.20.Cv}

\baselineskip=18pt
\section{Introduction}
The pioneering investigations of mathematical topics
related to the the black hole equilibrium states
were attributed to Israel \cite{isr},
M\"uller zum Hagen {\it et al.} \cite{mil73} and Robinson \cite{rob77}.
Bunting and Masood-ul-Alam \cite{bun} proposed 
the alternative proof of the uniqueness of black hole solutions.
Then, the method was strengthened to the Einstein-Maxwell (EM) black holes
\cite{ru,ma1}. Heusler \cite{he1} comprised the 
magnetically charged Reissner-Nordstr\"om (RN) 
solution and static Einstein-${\sigma}$-model case
\cite{he93}.
The classification of static of 
vacuum black holes was finished in \cite{chr99a}, where
the condition of non-degeneracy of the 
event horizon was removed and it was shown that Schwarzschild black hole exhausted
the family of all appropriately regular black hole spacetimes.
In Ref.\cite{chr99b} it was
revealed 
that RN solution comprised the family of regular
black hole spacetimes under the restrictive condition that all
degenerate components of black hole horizon carried a charge of the same sign.
\par
The problem of stationary
and axisymmetric black hole spacetimes being the solution of vacuum 
Einstein equations was considered in Refs.\cite{car73,car87,rob75,wal71},
while 
the systematic way of obtaining the desire results in
electromagnetic case was provided by Mazur 
\cite{maz} and Bunting \cite{bun}.
For a review of the uniqueness of black hole
solutions story see \cite{book} and references therein. 
\par
In the recent years there was a considerable resurgence of mathematical 
works concerning black hole equilibrium state in the low-energy string
theory. The uniqueness of the black hole solutions in dilaton gravity was  proved in 
works
\cite{mas93,gur95,mar01}, while the uniqueness of the static
dilaton $U(1)^2$
black holes being the solution of $N = 4, d = 4$ supergravity
was provided in \cite{rog99}. The extension of the proof to the theory
to allow for the inclusion of 
$U(1)^{N}$ static dilaton black holes was established in Ref.\cite{rog02}.
\par
The latest development of string theory 
as well as the possibility that the weak scale is the fundamental
scale of nature and the Planck scale is to be derived from it \cite{gva}
trigger the interests in higher
dimensional black hole solutions. 
The so-called
TeV gravity attracts attention to higher dimensional black hole 
which may be produced in high energy experiments \cite{gid}. It was revealed 
that
the five-dimensional stationary vacuum black hole are not unique. Myers-Perry
\cite{mye86} solution generalized the Kerr solution to arbitrary dimension, while Emparan
{\it et al.} \cite{emp02} revealed a counterexample showing that a five-dimensional
rotating black hole ring solution with the same angular momentum and mass
but the horizon of which was homeomorphic to $S^{2} \times S^{1}$.
The uniqueness theorem for $n$-dimensional
Schwarzschild-Tangherlini
black hole was provided by
Gibbons {\it et al.} \cite{gib02} and
for asymptotically flat
static $n$-dimensional
charged (dilaton) black holes was presented in Ref.\cite{gib02b}.
The uniqueness theorem for self-gravitating nonlinear $\sigma$-models
in higher dimensional spacetime was obtained in \cite{rog02a}.
As far as the stationary $n$-dimensional black hole uniqueness theorem is concerned,
the proof for $N = 1,~d = 5$ supersymmetric black holes was given in
\cite{rea02}.
\par
In this issue our strategy will be to modify the considerations
of Ruback \cite{rub88} and Chru\'sciel \cite{chr99b} in order to comprise
the problem of the uniqueness of the charged $n$-dimensional static 
black hole solutions containing an asymptotically flat spacelike 
hypersurface with compact interior and with both degenerate and
non-degenerate components of the event horizons.
In the attitude proposed by 
Ruback \cite{rub88} the metric induced on the hypersurface
orthogonal to the asymptotically timelike Killing vector field was considered.
In our case we shall
consider the {\it orbit space }metric \cite{chr99a} (defined below)
on the spacelike hypersurface $\Sigma$.
The key considerations will be held near both degenerate and non-degenerate 
components of $\p \Sigma$. 
Our key result will be the proof that a static higher dimensional charged black hole
spacetime containing an asymptotically flat spacelike hypersurface with
compact interior and with both degenerate and non-degenerate components of the 
event horizon exists, subject to the additional assumption that all
degenerate components of the horizon should have charges of the same signs.

\section{Higher dimensional Einstein-Maxwell system}
We shall consider the $n$-dimensional Einstein-Maxwell system described by the following
action:
\be
I = \int d^n x \sqrt{-{\hat g}} \bigg[ {}^{(n)}R - 
F_{\mu \nu}F^{\mu \nu}
\bigg],
\label{act}
\ee
where ${\hat g}_{ij}$ is $n$-dimensional metric tensor,
$F_{\mu \nu}$ is the Maxwell field.
The metric of $n$-dimensional static spacetime with
asymptotically timelike Killing vector field 
$k_{\alpha} = \big({\p \over \p t }\big)_{\alpha}$
and $\tV^{2} = - k_{\mu}k^{\mu}$ has the following form:
\be
ds^2 = - \tV^2 dt^2 + h_{i j}dx^{i}dx^{j},
\ee
where $\tV$ and $h_{i j}$
are independent of the $t$-coordinate as the quantities
of the hypersurface $\Sigma$ of constant $t$. 
The hypersurface $\Sigma$ is a connected and simply connected spacelike
hypersurface, $\bar \Sigma$ denotes the closure of it. The topological
boundary of $\Sigma$, ~$\p \Sigma = \bar \Sigma \backslash \Sigma$ is a nonempty
topological manifold with $h_{ij}k^{i}k^{j} = 0$ on $\p \Sigma$. For the static
metric the electromagnetic potential will be of the form $A_{0} = \phi dt$.
\par
We treat first the problem of the {\it orbit space}. Namely,
as in Ref.\cite{chr99a} we shall consider a point $p$ of the manifold such that
$h_{ij}k^{i}k^{j} \ne 0$ and perform a decomposition
\be
T_{p}M = {\cal K}(k_{\mu}) \oplus {\cal K}^{\perp},
\ee
where ${\cal K}(k_{\mu})$ is the vector space spanned by the Killing
vector field $k_{\mu}(p)$ and ${\cal K}^{\perp}$ is the space orthogonal to the
vector space spanned by $k_{\mu}$. We define the {\it orbit space}
metric $g_{ij}(p)$ in the form
\be
g(X, Z) = h(X_{\perp}, Z_{\perp}).
\ee
The metric $g_{ij}$ should not be identified with the metric on the {\it space of orbits}
because one is not assuming any regularity properties of that space (in particular
we do not even assume that the {\it space of orbits} is a differentiable manifold). One has
then
\be
Y_{\perp} = Y - {h(X, Y) \over h(X, X)}X,
\ee
so we arrive at the following expression for the metric tensor $g_{ij}$
\be
g(Y, Z) = h(Y, Z) - {h(X, Y) h(X, Z) \over h(X, X)}.
\ee
As was proved in Ref.\cite{chr99a} (lemma 5.1) if one had to do with the static spacetime 
and if we supposed
that $(h_{ij}, \tV)$ satisfied the same coordinate independent system
of equations, then the {\it orbit space} metric $g_{ij}$ with the function $V$ (where
$V^2$ was the square of the norm of $k_{\mu}$ on $\Sigma$) satisfied the same 
system of equations. This theorem enables us to write the equations of motion in the 
following form:
\ben
{}^{(g)}\na_{i} {}^{(g)}\na^{i} V &=& 
{C^2 \over V} {}^{(g)}\na_{i}\phi {}^{(g)}\na^{i} \phi, \\ 
{}^{(g)}\na_{i} {}^{(g)}\na^{i} \phi &=&  {1 \over V} 
{}^{(g)}\na_{i} \phi {}^{(g)}\na^{i} V, \\ 
{}^{(n - 1)}R_{ij}(g) - {1 \over V}{}^{(g)}\na_{i} {}^{(g)}\na_{j} V &=&
- {2 \over V^2} {}^{(g)}\na_{i}\phi {}^{(g)}\na_{j} \phi
+ {2 {}^{(g)}\na_{i} \phi {}^{(g)}\na^{i} \phi \over (n - 2) V^2} g_{ij},
\een
where we have denoted by $C^2 = 2(n - 3)/(n - 2)$, while $\phi$ is the
electrostatic potential. The covariant derivative 
with respect to 
$g_{ij}$ is denoted by ${}^{(g)}\na$,
while ${}^{(n - 1)}R_{ij}(g)$ is the Ricci tensor defined on 
the hypersurface $\Sigma$.
\\
Let us assume further
that we take into account the asymptotically flat spacetime,
i.e, the spacetime  
contains a data set
$(\Sigma_{end}, g_{ij}, K_{ij})$ with gauge fields such that 
$\Sigma_{end}$ is diffeomorphic to ${\bf R}^3$ minus a ball and the 
following asymptotic conditions are fulfilled:
\ben
\vert g_{ij}  - \delta_{ij} \vert + r \vert \p_{a}g_{ij} \vert
+ ... + r^k \vert \p_{a_{1}...a_{k}}g_{ij} \vert +
r \vert K_{ij} \vert + ... + r^k \vert \p_{a_{1}...a_{k}}K_{ij} \vert
\le {\cal O}\bigg( {1\over r} \bigg), \\
\vert F_{\alpha \beta} \vert + r \vert \p_{a} F_{\alpha \beta} \vert
+ ... + r^k \vert \p_{a_{1}...a_{k}}F_{\alpha \beta} \vert
\le {\cal O}\bigg( {1 \over r^2} \bigg).
\een
Consequently, under the above assumptions,
there is
a standard coordinates
system in which we have the usual asymptotic expansion
\be
V = 1 - {\mu \over r^{n - 3}} + {\cal O}\bigg( {1 \over  r^{n - 2}} \bigg),
\ee
and accordingly  for the metric tensor
\be
g_{ij} = \bigg( 1 + {2 \over (n - 3)}{\mu \over r^{n - 3}} \bigg) \delta_{ij} +
{\cal O} \bigg( {1 \over r^{n - 2}} \bigg),
\ee
and for the electrostatic potential
\be
\phi = {Q/C \over r^{n - 3}} + {\cal O} \bigg( {1 \over r^{n - 2}} \bigg),
\ee
where $\mu$ is the ADM mass seen by the observer from the infinity, 
$Q$ is the electric charge while
$r^2 = x_{i}x^{i}$.
\par
Our next step will be the analysis
of the behaviour of $g_{ij}$ and $\phi$ near $\p \Sigma$.
By the theorems derived by Vishweshwara \cite{vis68} and Carter \cite{car69}
one obtains that $\p \Sigma$ has to be a subset of a Killing horizon.
On the other hand, the Killing horizon is a smooth manifold.
As it is well known $\phi$ is constant on any connected components
of the Killing horizon and it implies its constancy on $\p \Sigma$.
In what follows, we shall call
a connected $S$ component of $\p \Sigma$
degenerate and non-degenerate
when $S$ intersects
a degenerate or non-degenerate Killing horizon, respectively. 
If necessary we can deform slightly the hypersurface $\Sigma$
in spacetime and we ensure that $\p \Sigma$ is a smooth submanifold both
of $\bar \Sigma$ and of $M$ near degenerate horizons.
As far as the non-degenerate horizon is concerned,
$\p \Sigma$ will not be a smooth submanifold of $M$. It is caused by the
existence of points on $\p \Sigma$ at which the Killing vector field
$k_{\mu}$ vanishes. As was proved in Ref.\cite{chr99a} one could equip
$\bar \Sigma$ with a differentiable structure so that $\p \Sigma$
was a smooth submanifold of $\bar \Sigma$. The boundary $\p \Sigma$
with the differentiable structure is a totally geodesic
boundary of $(\Sigma, g_{ij})$ across which $g_{ij}$ and the 
electric potential $\phi$ can be smoothly extented across $S$ when
a doubling of $\Sigma$ across $S$ is done.
\\
Based on the observation presented by Ruback \cite{rub88} let us consider now the function
\be
F_{\pm} = V^2 - \bigg( 1 \pm C\phi \Bigg)^2.
\ee
Having in mind equation of motion, by the direct calculations one can show that
$F_{\pm}$ is harmonic in the metric $V^{-2}g_{ij}$.
$F_{\pm}$ tend to zero as one approaches the asymptotically flat regions.
On the other hand,
on every components of $\p \Sigma$ one obtains $F_{\pm} \le 0$.
\par
First, let us suppose that 
$F_{-} = 0$ on all components of $\p\Sigma$. Next by means 
of the maximum principle one has that $F_{\pm} \equiv 0$ on $\bar \Sigma$.
In the case for which $F_{+}$ and $F_{-}$ are negative
somewhere on $\p \Sigma$, from maximum
principle we get that $F_{\pm} < 0$, provided that
\be
V^2 < \bigg( 1 - C\phi \bigg)^2 , \qquad V^2 < \bigg( 1 + C\phi \bigg)^2,
\label{vv1}
\ee 
on $\Sigma$. By hypothesis $V$ has no zeros on $\Sigma$
and having in mind (\ref{vv1}) it shows
that $\bigg( 1 - C\phi \bigg)$ and $\bigg( 1 + C\phi \bigg)$ have no zeros there.
Both $(1 - C\phi)$ and $(1 + C\phi)$ tends to $1$ at infinity which in turn implies that
$ -1 < C \phi < 1$ on $\Sigma$. The above equations implies that
$0 < V < \min \bigg( 1 + C\phi; 1 - C\phi \bigg) = 1 - C \mid \phi \mid$ on $\Sigma$
and we reach to the conclusion that
\be
0 \le V + C \mid \phi \mid \le 1.
\label{vv}
\ee
The inequalities are strict except when the metric is locally Majumdar-Papapetrou
(MP) \cite{mye87}.
The right inequality is strict on non-degenerate horizons.
\par
If $\mu = Q$ we have that $F_{+} = {\cal O}\bigg( {1 \over r^2} \bigg)$
 and $F_{+} \equiv 0$
follows from the harmonicity of $F_{+}$ in the metric $V^{-2}g_{ij}$ and the 
asymptotic strong maximum principle. Then, using the above arguments 
(as in proving Eq.(\ref{vv})) one reaches to the conclusion that the metric is locally
MP.
The case $\mu = - Q$ follows by providing the similar arguments as in the case
of $F_{+}$.
\par
Now we shall consider the case for which $\mu \ge \mid Q \mid$. 
In this stage of the proof we shall produce a data set to which one can apply 
considerations proposed by Rubback \cite{rub88} and Chru\'sciel \cite{chr99b}.
Namely,
we take into account two 
copies of the hypersurface $\Sigma_{+}$ and $\Sigma_{-}$ and define metric 
and the electric field $\hhE_{\alpha \pm}$ on them
\be 
\hhE_{\alpha \pm} = {1 \over \sqrt{2} \big( 1 + C \phi \pm V \big)}
\bigg[ \sqrt{2}~ {}^{(g)}\na_{\alpha}\phi \big(  1 + C \phi \big) -
\sqrt{{n-2 \over n - 3}} V {}^{(g)}\na_{\alpha} V \bigg],
\label{pme}
\ee
\be
\tg_{ij \pm} = \bigg( {1 + C \phi \pm V \over 2} \bigg)^{1\over n - 3} g_{ij}.
\label{met}
\ee
By $\p_{nodeg}\Sigma$ we shall denote all these components of the boundary of 
$\Sigma$ which correspond to non-degenerate components of the event
horizons of the black hole.\\
If $\p_{nodeg}\Sigma \ne 0$ we can paste $\bar \Sigma_{+} = \Sigma_{+}
\cup \p_{nodeg}\Sigma$ and $\bar \Sigma_{-} = \Sigma_{-}
\cup \p_{nodeg}\Sigma$ by indentifying $\p_{nodeg}\Sigma$ from $\bar \Sigma_{+}$
with $\p_{nodeg}\Sigma$ from $\bar \Sigma_{-}$ using the identity map.
Then, we obtain
$\hat \Sigma = \Sigma_{+} \cup \Sigma_{+} \cup \p_{nodeg}\Sigma$.
The metric defined on $\Sigma_{+} \cup \Sigma_{-}$ can be extended by
continuity to smooth metric on $\hat \Sigma$, similarly
this can be done with the electric field. Thus, we set that
\be
{\hat g}_{ij}\mid_{\Sigma_{+}} = \tg_{ij +}  \qquad 
{\hat g}_{ij}\mid_{\Sigma_{-}} = \tg_{ij -}.
\ee
and consequently
\be
{\hat E} \mid_{\Sigma_{+}}
= {\hat E}_{+}, \qquad 
{\hat E} \mid_{\Sigma_{-}}
= {\hat E}_{-}.
\ee
In the case of $\p_{nodeg}\Sigma = 0$, one has that ${\hat \Sigma} = \Sigma$,~
$\hat g_{ij} = \tg_{ij}$ and $\hat E = \hat E_{+}$.
\par
A tedious but simple calculations can envisage the following fact:
\ben \label{tsym}
{}^{(n - 1)}\tR_{ij}(\tg) = 2 \hhE_{\alpha}\hhE^{\alpha }, \\ \label{tsym1}
{}^{(\tg)}\na_{\alpha} \hhE^{\alpha } = 0,
\een 
where ${}^{(\tg)}\na_{\alpha}$ is the covariant derivative with respect to the metric
$\tg_{ij}$. 
Eqs.(\ref{tsym}) and (\ref{tsym1})
consist (as was observed for the first time in Ref.\cite{rub88}) the
constraint relations for a time symmetric problem.
\par
Using the asymptotical behaviour of $V$ and $\phi$ 
it could be observed
that as if we approach
$\Sigma_{+}$ the line element $ds^2_{+}$ could be expressed as
\be
ds^2_{+} = \bigg( 1 + {\mu + Q \over (n - 3) r^{n-1}} \bigg)
dx_{i} dx^{i} + {\cal O} \bigg( {1 \over r^{n - 2}} \bigg).
\label{adm}
\ee
From Eq.(\ref{adm}) it follows in particular that,
the ADM mass is equal to $M_{ADM} = {\mu + Q \over 2}$. It can be verified
that,
the electric field is given by
\be
\hhE_{\alpha} = {1 \over 2 \sqrt{2}} \bigg[
- \sqrt{(n - 3) (n - 2)} \bigg( {\mu + Q \over r^{n - 2}} \bigg) \bigg]
+ {\cal O} \bigg( {1 \over r^{n - 2}} \bigg).
\ee
On $\Sigma_{-}$ the metric is as follows:
\be
ds^2_{-} =  {1 \over 2^{2 \beta}} \bigg[
{\mu + Q \over r^{n - 3}} \bigg]^{2 \beta} dx_{i} dx^{i}
 + {\cal O} \bigg( {1 \over r^{n - 2}} \bigg),
\ee
where $\beta = {1 \over n - 3}$.\\
Further on, one can introduce the coordinates $x_{I}$, where $I = 1,...,N-2$,
so that the metric tensor $g_{ij}$ can be rewritten in the form
\be
g_{ij} = {1 \over W^2}dV^2 + \ga_{IJ} dx_{I} dx^{J}.
\ee
From the exact form of $\hhE_{\alpha \pm}$ Eq.(\ref{pme}) 
and having in mind Eqs.(\ref{tsym}) and (\ref{tsym1})
satisfied by them one can deduce that
the gauge potential is a function of $V$, namely
$\phi(V)$. This fact in turn enables us to rewrite equations of motion as follows:
\ben \label{rrr}
{\p \phi \over \p V} = {V \over W \sqrt{\ga}} {\p \over \p V} \bigg(
W \sqrt{\ga} {\p \phi \over \p V} \bigg), \\
C^2 \bigg ( {\p \phi \over \p V} \bigg)^2 = {V \over W \sqrt{\ga}} 
{\p \over \p V} \bigg( W \sqrt{\ga} \bigg),
\label{rrr1}
\een
where $\ga = det \ga_{ij}$.\\
Accordingly to Eqs.(\ref{rrr}) and (\ref{rrr1}) we obtain
\be
{\p \phi \over \p V} - C^2 \bigg( {\p \phi \over \p V} \bigg)^3 = 
V {\p^2 \phi \over \p V^2}.
\ee
From the above relation it follows in particular that
\be
{\p \phi \over \p V} = { a V \over \sqrt{1 + a^2 C^2 V^2}},
\label{ff}
\ee
where $a$ is an integration constant.
Using the asymptotic behaviour of $V$ we get the following expression:
\be
\lim\limits_{V \rightarrow 1} {\p \phi \over \p V} = - {(n - 3)Q \over \mu}.
\ee
Integrating Eq.(\ref{ff}) and taking the limit
$\lim\limits_{V \rightarrow 1} \phi = 0$ we arrive at the following 
expression for $\phi(V)$:
\be
\phi(V) = {\mu - \sqrt{\mu^2 + Q^2 (n -3)^2 (V^2 - 1)} \over
(n - 3)C Q}.
\ee
\par
By virtue of Eq. (\ref{tsym}) 
one can see that the Ricci tensor ${}^{(n - 1)}\tR_{ij}(\tg)$
on the hypersurface is manifestly non-negative. Furthermore the asymptotic
behavoiur of the metric $\tg_{ij +}$ on $\Sigma_{+}$ becomes
\be
\tg_{ij +} = \delta_{ij} + {\cal O}\bigg( {1 \over  r^{n - 2}} \bigg).
\label{gp}
\ee
While on $\Sigma_{-}$ it reduces to
\be
\tg_{ij -} = \bigg[ \bigg( {\mu + Q \over 2}\bigg)^{1 \over n - 3} 
{1 \over r} \bigg] \delta_{ij} +
{\cal O}\bigg( {1 \over  r^2} \bigg).
\label{gn}
\ee
It follows directly from
Eq.(\ref{gp}) that the total mass on $\hat \Sigma$ vanishes.
Thus, as a consequence of the positive mass theorem
\cite{posth}, the manifold $\hat \Sigma$ 
is isometric to flat manifold. One 
 can rewrite $g_{ij}$ in a 
conformally flat form \cite{gib02}
\be
g_{ij} = {\cal U}^{1 \over n-3} \delta_{ij},
\label{gg}
\ee
where we have defined a smooth function on $(\hat \Sigma, \delta_{ij})$,
namely ${\cal U} = {2 \over 1 + V + C\phi}$.
One can show that the
Einstein-Maxwell equations of motion reduces
to the Laplace equation on the $(n - 1)$ Euclidean manifold 
$
\na_{i}\na^{i}{\cal U} = 0,
$
where $\na$ is the connection on a flat manifold. 
Having in mind the above we can adopt for the metric $\delta_{ij}$
in the flat base space the following metric:
\be 
\delta_{ij} dx^{i}dx^{j} = \trho^{2} d{\cal U}^2 + \th_{AB}dx^{a}dx^{B}.
\ee
First we shall consider the case of the single horizon.
The event horizon is located at ${\cal U} = 2$ and
one can show that the embedding of $\cal H$ into the Euclidean
$(n-1)$ space is totally umbilical \cite{kob69}. This embedding must be 
hyperspherical, i.e., each of the connected components of the horizon $\cal H$
is a geometric sphere with a certain radius determined by the value of
$\rho \mid_{\cal H}$,
where $\rho$ is the coordinate which can be introduced on $\Sigma$
as follows:
$$g_{ij}dx^{i}dx^{j} = \rho^2 dV^2 + h_{AB}dx^{A}dx^{B}.$$
One can always locate one 
connected component of the horizon at $r = r_{0}$ surface without loss of generality.
Thus, we have to do with a boundary value problem for the Laplace equation 
on the base space $\Omega = E^{n-1}/B^{n-1}$ with the
Dirichlet boundary condition ${\cal U} \mid_{\cal H} = 2$ and the asymptotic
decay condition ${\cal U} = 1 + {\cal O} \bigg( {1 \over r^{n-3}} \bigg)$.
Suppose further that
${\cal U}_{1}$ and ${\cal U}_{2}$ be two solutions of the boundary value problem.
By means of the Green identity and integration over the volume element
we reach to the following expression:
\be
\bigg( \int_{r \rightarrow \infty} - \int_{\cal H} \bigg) 
\bigg( {\cal U}_{1} - {\cal U}_{2} \bigg) {\p \over \p r}
\bigg( {\cal U}_{1} - {\cal U}_{2} \bigg) dS = \int_{\Omega}
\mid \na \bigg( {\cal U}_{1} - {\cal U}_{2} \bigg) \mid^{2} d\Omega.
\ee
Because of the boundary condition the left-hand side vanishes, and we draw
a conclusion 
that two solutions must be identical.
\par
The case of not single horizon can be treated as in Ref.\cite{gib02,gib02b}.
Namely, one should consider the evolution level surface in Euclidean space. 
From the Gauss equation in Euclidean space we shall obtain the evolution equation for 
shear
$\sigma_{AB}$. Making use of the harmonicity of $\cal U$, i.e., $\na^2 {\cal U} = 0$, 
we can draw a conclusion that
\be
\sigma_{AB} = 0, \qquad \hat {\cal D}_{A} \rho = 0, \qquad \hat {\cal D}_{A} k = 0,
\ee
where $\hat {\cal D}_{A}$ denotes the covariant derivative on each level set of $V$,
$k_{AB}$ is the second fundamental form of the level set. This implies in turn that 
each level surface of $\cal U$ is totally umbilic and hence spherically symmetric.
\par
As was mentioned in Ref.\cite{gib02,gib02b} this is the local result, since one consider 
only the region without saddle points of the harmonic function $\cal U$. 
In order to achieve the global result one should take into account the
assumption about analyticity.
\par
Now we turn to the problem of a charge of the connected components of black hole
event horizon
and its generalization to the non-connected case. 
In four-dimensions
Heusler \cite{heu97} showed that if all horizons were degenerate and $Q_{i}Q_{j} \ge 0$,
where $Q_{i}$  was the charge of the adequate connected component of black hole, 
then the black hole was a standard MP black hole. 
One should has in mind that a standard connected MP
black hole is an extreme RN one \cite{mye87}.
\par
First,
if we suppose that $S_{a}$, where $a = 1,2$, is a 
connected components of $\p \Sigma$ such that the 
electrical
potentials of the horizons
$\phi_{a} = \phi \mid_{S_{a}}$ imply
\be
\phi_{1} = \inf_{\bar \Sigma} < 0, \qquad \phi_{2} = \sup_{\bar \Sigma} > 0.
\ee
Then as was shown in \cite{chr99b} $Q_{a}$ of the $S_{a}$ are non-vanishing and have 
the opposite signs.
It can be verified by writing out the expression for
the charge $Q_{a}$ 
\be
Q_{a} = - \lim \limits_{i \rightarrow \infty} \int_{S_{a,i}}
{C \na^{i} \phi \over V}dS_{i}.
\ee
and using the divergence theorem and maximum principle (see \cite{chr99b} for the details).
\par
By exactly similar arguments as presented in work \cite{chr99b}
we can treat the case of a non-connected $\p \Sigma$. Finally, the 
above considerations enable us to formulate the main conclusion
of our work.\\
{\bf Theorem:}\\
Let us consider a static solution to the $n$-dimensional Einstein-Maxwell
equations of motion with an asymptotically timelike Killing vector field $k_{\mu}$.
Suppose further that the manifold under consideration consists 
of a connected and simply connected
spacelike hypersurface $\Sigma$ to which $k_{\mu}$ is orthogonal. The topological
boundary $\p \Sigma$ of $\Sigma$ is a nonempty topological manifold with
$h_{ij}k^{i}k^{j} = 0$ on $\p \Sigma$.
Then, one arrives at the following conclusions:

\begin{enumerate}
\item{If $\p \Sigma$ is connected, then there exist a neighbourhood of $\Sigma$ which is
diffeomorphic to an open subset of $n$-dimensional Reissner-Nordstr\"om 
spacetime (extreme or non-extreme).}
\item{If $\p \Sigma$ is not connected ad moreover we have 
fulfilled the following inequality:
\be
\forall_{i, j}~ Q_{i}Q_{j} \ge 0,
\ee
where $Q_{i}$ is the charge of the adequate component of 
$\p \Sigma$, that intersects the degenerate horizon, then there is
an open neighbourhood of $\Sigma$ which is diffeomorphic to $n$-dimensional
Majumdar-Papapetrou spacetime.}
\end{enumerate}                      

\vspace{0.5cm}
\noindent
{\bf Acknowledgements:}\\
MR was supported in part by KBN grant 2 P03B 124 24.



\end{document}